\documentclass[showpacs,preprintnumbers,amsmath,amssymb,APSl,prd,twocolumn,nofootinbib,superscriptaddress]{revtex4-2}

\usepackage{bm}
\usepackage{mathrsfs}
\usepackage{xcolor,color,graphicx,graphics}
\usepackage{colortbl}  
\usepackage{epsfig,subfigure}
\usepackage{latexsym,amssymb,amsmath,amsfonts} 
\usepackage[english]{babel} 
\usepackage[OT1]{fontenc}
\usepackage[latin1]{inputenc}
\usepackage{makeidx}
\usepackage{hyperref}
\usepackage{color,graphicx,graphics,wrapfig,epsf}
\usepackage{hyperref}
\hypersetup{colorlinks=true, linkcolor=blue, citecolor=green}
\usepackage{makecell}
\usepackage{multirow}
\def\imo{i}

\def\K{{\cal K}}
\def\Order#1{{\cal O}\left(#1\right)}
\usepackage{appendix}
\usepackage{orcidlink}

\usepackage{lipsum}     
\usepackage{mathtools}

\usepackage{latexsym}
\usepackage{orcidlink}
\usepackage{enumerate}

\definecolor{red}{rgb}{1,0,0}

\def\+{^\dagger}

\def\<{\leftarrow}
\def\>{\rightarrow}

\def\({\left(}
\def\){\right)}

\def\K{{\cal K}}

\newcommand{\bi}{\begin{itemize}} 				\newcommand{\ei}{\end{itemize}}
\newcommand{\benu}{\begin{enumerate}} 		\newcommand{\enu}{\end{enumerate}}
\newcommand{\bd}{\begin{dinglist}{0}}     \newcommand{\ed}{\end{dinglist}}
\newcommand{\bfig}{\begin{figure}[htbp]}  \newcommand{\efig}{\end{figure}}
        			
\newcommand{\bc}{\begin{center}} 				  \newcommand{\ec}{\end{center}}
\newcommand{\be}{\begin{equation}} 				\newcommand{\ee}{\end{equation}}
\newcommand{\bsub}{\begin{subequations}}  \newcommand{\esub}{\end{subequations}}
\newcommand{\ben}{\begin{eqnarray}} 			\newcommand{\een}{\end{eqnarray}}
\newcommand{\ba}[1]{\begin{array}{#1}} 		\newcommand{\ea}{\end{array}}
\newcommand{\bea}{\begin{equation}\begin{array}{rcl}}
\newcommand{\eea}{\end{array}\end{equation}}

\usepackage{soul}

\begin{document}
\title{Black hole physics within $f(R)$ gravity: Quasi-normal spectra and greybody factors}

\author{\'Angel Rinc{\'o}n 
        \orcidlink{0000-0001-8069-9162}} 
        \email{angel.rincon@physics.slu.cz}
        \affiliation{Departamento de F{\'i}sica, Universidad del B{\'i}o-B{\'i}o,
Casilla 5-C, Concepci{\'o}n, Chile.}
\affiliation{Research Centre for Theoretical Physics and Astrophysics, Institute of Physics, Silesian University in Opava, Bezru\v{c}ovo~n\'am\v{e}st\'i 13, CZ-74601 Opava, 
        Czech Republic.}

\author{Grigoris Panotopoulos 
\orcidlink{0000-0002-7647-4072}} \email{grigorios.panotopoulos@ufrontera.cl}
\affiliation{Departamento de Ciencias F{\'i}sicas, Universidad de La Frontera, Casilla 54-D, 4811186 Temuco, Chile.}

\date{\today}

\begin{abstract}
We investigate several astrophysical motivated properties of a four-dimensional black hole solution in the framework of $f(R)$ gravity. The model is characterized by a single additional parameter, $\alpha$, which encodes nontrivial deviations from General Relativity. Using this black hole spacetime as the background geometry, we analyze:
(i) the quasinormal modes of massless scalar perturbations (employing several complementary methods), and
(ii) the greybody factors associated to the propagation of massless test scalar fields. 
Regarding quasinormal modes, we study the response of black holes to massless scalar perturbations using three independent approaches: the well-established sixth-order WKB semi-analytic method, analytic expressions, and the analytical expression in the eikonal limit. We examine the behavior of both the real and imaginary parts of the quasinormal frequencies as functions of the parameter $\alpha$, the overtone number $n$, and the multipole number $\ell$.
In addition to that, we compute the greybody factors within the WKB approximation for various combinations of the relevant parameters. In particular, we focus on the propagation of test massless scalar fields and investigate how the greybody factors depend on the multipole number $\ell$, and the free parameter $\alpha$. The impact of the aforementioned parameters on the absorption cross-section and the reflection and transmission coefficients is studied in detail.
\end{abstract}

\maketitle


\tableofcontents

\section{Introduction}\label{intro}

After Hawking's seminal papers where he showed that black holes (BHs) emit radiation from their horizon \cite{Hawking1,Hawking2}, BHs have become exciting objects and an excellent laboratory to study and understand quantum gravity. It is often said that the Hawking radiation is a black body radiation, thermal in nature,
characterized entirely by the hawking temperature $T_H$. This, however, is only partially true. The reason why is that the emitted particles feel an effective potential barrier in the exterior region. The potential barrier backscatters a part of the outgoing radiation back into the black hole \cite{Kanti}. The greybody factor (GBF), or else absorption cross section $\sigma_{abs}(\omega)$, is a frequency dependent factor that measures the modification of the original black body radiation. The total black hole emission rate is obtained by integrating the greybody factor $\sigma_\ell$ over all spectra. Moreover, if the greybody factor is a constant the black hole emission spectrum would be exactly that of a blackbody radiation. This is the non-triviality of the greybody factor which leads to deviations of blackbody emissions and the consequent greybody radiation \cite{Campuzano}.
Greybody factors are important both from the theoretical and from the experimental point of view \cite{Kanti}. From the theory side, they give us valuable information about the near horizon structure of black holes. From the experiment side, although the Hawking radiation has not been detected yet, is that greybody factors modify the spectrum in the region where most particles are produced. This could be essential in studying collider signatures of the evaporation of TeV mini black holes \cite{Casanova}.

Ideal BHs are supposed to be isolated objects. Realistic BHs of Nature, however, are in constant interaction with their environment. We may think, for instance, matter accretion onto a BH from its donor in binaries. When a BH is perturbed due to a certain interaction, the geometry of space-time undergoes damped oscillations. How a system responds to small perturbations as well as normal modes of oscillating systems have always been important topics in physics. Regarding BH physics in particular, the work of \cite{regge} long time ago marked the birth of BH perturbation theory, and later on it was extended by other people \cite{zerilli1,zerilli2,zerilli3,moncrief,teukolsky}. Nowadays the state-of-the art in BH physics and perturbations is nicely summarized in the comprehensive review of Chandrasekhar's monograph \cite{monograph}. The information on how a given BH relaxes after the perturbation has been applied is encoded into the quasi-normal (QN) frequencies. The latter are complex numbers, with a non-vanishing imaginary part, that depend on the details of the background geometry as well as the spin of the propagating field at hand (scalar, Dirac, vector (electromagnetic), tensor (gravitational)), but they do not depend on the initial conditions. Therefore, QN modes (QNMs) carry unique information about black hole physics. Black hole perturbation theory and QNMs of black holes are relevant during the ringdown phase of binaries, in which after the merging of two black holes a new, distorted object is formed, while at the same time the geometry of space-time undergoes damped oscillations due to the emission of gravitational waves. For a review on QNMs of BHs see e.g. \cite{Berti:2009kk, Konoplya:2011qq}.

In order to formulate a self-consistent theory of gravity, it is essential to incorporate quantum features in such a way that both large-scale phenomena and small-scale (microscopic) physics are simultaneously described, avoiding infinities, singularities, ghosts and other pathologies. In this context, a viable theory of gravity should, for instance, address the singularity problem inherent in General Relativity (GR). Such singularities are typically characterized by the unbounded growth of curvature invariants, rendering GR unpredictable in the most extreme regimes of the Universe.

Motivated by the need to reconcile GR with Quantum Mechanics (QM) and to address outstanding cosmological challenges, such as dark energy and inflation, a wide variety of alternative theories of gravity have been proposed as promising extensions of GR. Among these are the Scalar-Tensor theories, which introduce an additional scalar degree of freedom coupled to the metric tensor. These models allow for dynamical variations of the effective gravitational coupling and provide natural mechanisms for cosmic acceleration. A prototypical example is Brans-Dicke theory \cite{Brans:1961sx,Brans:1962zz}, while more general frameworks, such as Horndeski gravity \cite{Horndeski:1974wa,Babichev:2016rlq,Babichev:2017guv}, extend this class by preserving second-order field equations while accommodating richer gravitational dynamics.

An alternative route is provided by teleparallel gravity \cite{Aldrovandi:2013wha,Hohmann:2022mlc}, which reformulates gravity in terms of torsion rather than curvature. While its simplest formulation is dynamically equivalent to GR, teleparallel gravity admits natural extensions, such as $f(T)$ gravity, that have been extensively explored in cosmological applications \cite{Cai:2015emx}. Similarly, Gauss-Bonnet gravity \cite{Fernandes:2022zrq,Garcia-Aspeitia:2020uwq} incorporates higher order curvature invariants, notably the Gauss-Bonnet term, which becomes dynamically relevant in higher-dimensional spacetimes \cite{Lovelock:1971yv} or suitably modified four-dimensional formulations including a scalar field. Such theories may offer resolutions to spacetime singularities or explanations for early-Universe dynamics.

A particularly prominent quantum gravity inspired framework is Asymptotically Safe Gravity, which provides a potential ultraviolet completion of GR through the existence of a non-Gaussian fixed point in the renormalization group flow. This mechanism ensures predictivity at high energies without relying on perturbative renormalizability. Originally proposed by Weinberg \cite{Weinberg1978,2010grae.book.....H}, asymptotic safety has received substantial support from functional renormalization group studies in both pure gravity and matter-coupled systems, with applications ranging from cosmology and black hole physics to {\bf{small scales}} \cite{Bonanno:2017pkg,Falls:2010he,Bosma:2019aiu}. More recent investigations have further explored nonlocal effects, higher-derivative truncations, and issues of background independence.

Closely related to asymptotically safe gravity is the scale-dependent gravity approach, in which gravitational couplings, such as Newton's constant and the cosmological constant, are promoted to scale-dependent quantities that evolve with the energy scale due to quantum effects \cite{Koch:2010nn,Koch:2014joa}. This framework has been successfully applied to a wide range of gravitational systems, including cosmological models \cite{Alvarez:2022mlf,Alvarez:2022wef,Panotopoulos:2021heb}, black holes \cite{Koch:2016uso,Rincon:2017goj,Rincon:2018dsq}, wormholes \cite{Contreras:2018swc}, and compact objects \cite{Panotopoulos:2021obe}. In many cases, scale dependence offers novel mechanisms for addressing long-standing issues such as dark energy, spacetime singularities, and the behavior of gravity at high energies.

Black holes provide an exceptional arena for probing not only the classical but also the quantum aspects of gravity. Remarkably, starting from relatively ordinary initial conditions, such as the gravitational collapse of a massive star, Nature produces spacetime geometries capable of amplifying short-distance quantum fluctuations to macroscopic scales.

Given their fundamental importance, BHs can be regarded as natural ''laboratories'' for testing gravitational theories. Consequently, irrespectively of the specific theoretical framework, any viable black hole solution should lead to testable predictions for observable phenomena. In this context, beyond resolving spacetime singularities and incorporating quantum effects, it is crucial to assess whether or not such solutions yield observable signatures in the region exterior to the event horizon.

Among the various observables currently accessible, only a few stand out as particularly promising. Notably, quasinormal modes (QNMs) and greybody factors (GBFs)  \cite{Pedrotti:2025idg} have emerged as leading probes in the present golden era of gravitational-wave Astronomy and multi-messenger Astrophysics.

Briefly, (i) QNMs correspond to the characteristic damped oscillations of a perturbed black hole as it relaxes toward its final stationary state. They dominate the ringdown phase of gravitational-wave signals and are highly sensitive to the underlying spacetime geometry; (ii) GBFs quantify the modification of the Hawking radiation spectrum observed at infinity, leading to deviations from an ideal blackbody distribution. These factors, which take values between 0 and 1, arise from the scattering of outgoing radiation by the effective gravitational potential barrier.

Given the importance of the GBFs and QNMs, it would be interesting to see what kind of spectra and absorption cross-sections are expected in the context of well-motivated alternative theories of gravity. The aim of this work is to investigate a specific four-dimensional BH solution within the framework of a concrete $f(R)$ gravity characterized by a single free parameter. We analyze how modifications of the spacetime geometry affect the QN mode spectrum, GBFs, and BH shadow. Our primary objective is to assess whether this solution provides meaningful insights into the nature of BHs in $f(R)$ gravity, and to evaluate its potential for observational tests and falsifiability.

In the present article we organize our work as follows: After this introductory section, we present the action, field equations and black hole solution in Section \ref{S:II}. In the third Section we briefly review QNMs of black holes for scalar perturbations. The WKB method and the eikonal limit are also discussed. In Section \ref{S:IV} the GBFs of black holes are presented, where the absorption cross section and decay rate are defined. The main numerical results are discussed in the fifth section. Finally, we summarize our work and conclude in Section \ref{S:VI}. We adopt the mostly positive metric signature, and we work in natural units where we set $c = 1$.

\section{Background and formalism} \label{S:II}

 \subsection{$f(R)$ gravity in arbitrary number of dimensions}\label{sec2}

In this subsection, we aim to summarize the essential ingredients needed to understand the framework of $f(R)$-gravity, an natural modification of GR in which we accept high order correction in terms of Ricci scalar, one of several promising approaches to modify GR maintaining the regime in which general relativity works well.
Thus, let us start by considering the standard classical action
\begin{equation}
S\,\equiv\,S_g+S_m,
\end{equation}
where $S_g$ is, in general, the $D$ dimensional gravitational action given by the expression:
\begin{equation}
S_g=\frac{1}{16 \pi G_D}\int \text{d}^{D}x\sqrt{-g}\,[R+f(R)],
\label{S_g}
\end{equation}
where the parameters involve have the usual meaning, i.e.: 
  i) $G_D \equiv M_D^{2-D}$ is the $D$ dimensional Newton's constant,  
 ii) $M_D$ is the corresponding Planck mass, 
iii) $g$ is the determinant of the metric $g_{AB}$, $(A,B=0, 1, ..., D-1)$, 
 iv) $R$ represents the scalar curvature and finally
  v) $R+f(R)$ is the function defining the theory to be studied.
For simplicity let us consider the corresponding Einstein-Hilbert action, including a non-vanishing cosmological constant, $\Lambda_D$, given by $f(R)=-(D-2)\Lambda_D$.
\\
The matter contribution $S_m$ defines, as usual, the energy-momentum tensor by the use of the following expression:
\begin{equation}
T^{AB}=-\frac{2}{\sqrt{-g}}\frac{\delta S_m}{\delta
g_{AB}}.
\end{equation}
Deriving the corresponding equation of motion (in the metric formalism):
\begin{align}
\begin{split}
 R_{AB}(1+f'(R)) - \frac{1}{2}(R+f(R))\,g_{AB}&
\\
+(\nabla_A \nabla_B-g_{AB}\Box)f'(R) &= 8\pi G_D T_{AB},
\label{Einsteins_eqns}
\end{split}
\end{align}
where $R_{AB}$ is the Ricci tensor and $\Box=\nabla_A \nabla^A$ with $\nabla$
the usual covariant derivative.
Thus for the vacuum Einstein-Hilbert action (with cosmological constant) we have the corresponding Einstein field equations:
\begin{eqnarray}
 R_{AB}-\frac{1}{2}R\,g_{AB}+\frac{D-2}{2}\Lambda_D g_{AB}=0,
\end{eqnarray}
where $R_{AB}=\Lambda_D g_{AB}$ and $R= D \Lambda_D$.
Notice that to get constant scalar curvature solutions $R\,=\,R_0$ the condition is
\begin{eqnarray}
 R_{AB}\,(1+f'(R))-\frac{1}{2}\,g_{AB}\,(R+f(R))\,=\,0.
\end{eqnarray}
Taking the trace in previous equation, $R_0$ must be a root of the equation:
\begin{eqnarray}
 2(1+f'(R_0))\,R_{0}-D\,(R_{0}+f(R_{0}))\,=\,0.
\label{root_R0}
\end{eqnarray}
A cosmological constant may be naturally defined as $\Lambda_D^{\text{eff}}\equiv R_{0}/D$ for this kind of solutions. Thus, 
for constant curvature solution $R=R_0$ with $1+f'(R_0)\neq 0$ we have:
\begin{eqnarray}
 R_{AB}=\frac{R_{0}+f(R_0)}{2(1+f'(R_0))}\,g_{AB},
\end{eqnarray}
On the other hand one can consider:
\begin{equation}
2R\,(1+f'(R))-D\,(R+f(R))\,=\,0,
\label{dif}
\end{equation}
as the differential equation that the function $f(R)$ must satisfy in order for the corresponding solution to accommodate an arbitrary constant curvature $R$. The solution to this differential equation is given by:
\begin{equation}
f(R)\,=\,a R^{D/2}-R,
\end{equation}
where $a$ is an arbitrary constant. 
In addition, notice that, as the Lagrangian becomes proportional to $a R^{D/2}$ which will have solutions of constant curvature for arbitrary $R$. 
In what follows, we will have to face the problem of establishing a clear criteria to relate solutions of the Einstein-Hilbert action with solutions of more general $f(R)$ gravities (i.e., beyond the constant curvature case).
Let $g_{AB}$ a solution of Einstein-Hilbert gravity with cosmological constant, i.e.:
\begin{eqnarray}
 R_{AB}-\frac{1}{2}R\,g_{AB}+\frac{D-2}{2}\Lambda_D g_{AB} = 8\pi G_D T_{AB}.
\label{LambdaCDM_eq}
\end{eqnarray}
Then $g_{AB}$ is also a solution of any $f(R)$ gravity, provided the following
equation
\begin{align}
\begin{split}
f'(R) R_{AB} &- \frac{1}{2}g_{AB}\left[f(R)+(D-2)\Lambda_D\right]
\\
&+ (\nabla_A\,\nabla_B-g_{AB}\Box)f'(R)=0,
\label{comp}
\end{split}
\end{align}
obtained from (\ref{Einsteins_eqns}) is fulfilled.
This formalism and key examples are discussed in detail in Ref.~\cite{delaCruz-Dombriz:2009pzc}, where the relevant equations are presented comprehensively. For instance, the reference covers the vacuum case ($T_{AB}=0$) with vanishing cosmological constant ($\Lambda_D=0$), as well as the case with $\Lambda_D=0$ and conformal matter ($T=T^A_A=0$). Both of these standard scenarios are summarized there.

\subsection{Concrete model: $f(R)=2 \alpha \sqrt{R}$}

In what follows, we investigate a spherically symmetric black hole solution in four-dimensional $f(R)$ gravity, where the functional form is chosen as $f(R) = 2 \alpha \sqrt{R}$, with $\alpha < 0$ being a dimensionful parameter that encodes deviations from General Relativity (GR). 
This model, inspired by square-root modifications to the Einstein-Hilbert action, aims to address issues such as small scales and quantum corrections while maintaining compatibility with GR limits, i.e., when $\alpha$ goes to zero. The inclusion of the square-root term $\sqrt{R}$ introduces non-linear effects that might, in principle, regularize the spacetime geometry at short distances, potentially avoiding the central singularity typical in GR black holes.
The modified action, in (1+3)-dimensions, acquire the following concrete form:
\begin{equation}
S[g_{\mu\nu}] = \int d^4x \sqrt{-g} \left[\frac{1}{2\kappa}\Bigl(R + 2 \alpha \sqrt{R} \Bigl)\right] + S_m,
\end{equation}
where the symbols retain their conventional meanings: 
i) $R$ is the Ricci scalar, 
ii) $\alpha$ is the parameter that controls the deviation from the Einstein-Hilbert action, and 
iii) $S_m$ represents the matter action. 
We work in units where $c = 1$ and $\kappa \equiv 8\pi G$.
Assuming a static and spherically symmetric spacetime, the line element takes the form  
\begin{equation}\label{m1}
\mathrm{d}s^{2} = -A(r)\,\mathrm{d}t^{2} + B(r)\,\mathrm{d}r^{2} + r^2\,\mathrm{d}\Omega^2,
\end{equation}  
where $\text{d}\Omega^2 \equiv \text{d}\theta^2 + \sin^2\theta\,\text{d}\phi^2$. For simplicity, we consider the case $A(r) = B(r)^{-1}$. 
This black hole previously found in~\cite{Clifton:2005aj,Sebastiani:2010kv}, is described by the metric functions  
\begin{equation} \label{metricelement}
A(r) = B(r)^{-1} = \frac{1}{2} + \frac{1}{3\alpha r} ,
\end{equation}
where $\alpha$ can be identified with the event horizon $r_H \equiv 2/(-3\alpha)$.
Notice that the metric function is very similar to the one corresponding to the Schwarzschild solution. Since the Ricci scalar is computed to be $R=1/r^2$, the spacetime is not flat. However, contrary to the Schwarzschild geometry, the solution considered here is not asymptotically flat as some components of the Riemann tensor grow with $r$ \footnote{We thank N.~Dimakis for his help on that issue.}. This issue can be addressed producing as a result a deficit angle, as we can show in brief. 
Firstly, note that when we rewrite the lapse function in terms of the event horizon, a general scale factor naturally emerges, suggesting that we should redefine the variables accordingly. Thus we have
\begin{equation} \label{metricelement2}
A(r) = B(r)^{-1} \equiv \frac{1}{2} \Bigg(1 - \frac{r_H}{r} \Bigg) ,
\end{equation}
The metric given in Eq.~\eqref{metricelement} can be recast by introducing $T \equiv t/\sqrt{2}$ and $x \equiv \sqrt{2} \ r$, inducing an angular deficit on $\Omega$, because $\mathrm{d}\tilde{\Omega}^2 \equiv (1/2) \mathrm{d}\Omega^2$ Thus, the line element becomes 
\begin{equation}\label{m2}
\mathrm{d}s^{2} = -A(x)\,\mathrm{d}T^{2} + B(x)\,\mathrm{d}x^{2} + x^2\,\mathrm{d}\tilde{\Omega}^2,
\end{equation}
the and the metric functions take the simplest form
\begin{equation}
A(x) = B(x)^{-1} \equiv 1 - \frac{x_H}{x},
\label{metricelement2}
\end{equation}
The event horizon, which defines the limit in which trapped surfaces are developed, can be obtained simply by finding the roots of $A(x_H) = 0$. 
In this case, it becomes obvious that the horizon is $x_H = 2\sqrt{2}/(-3\alpha)$.

\section{Quasinormal modes}\label{S:III}

We present the essential equations and theoretical background needed to compute quasinormal modes (QNMs). The analysis focuses on a Schwarzschild-like black hole discussed earlier in the context of $f(R)$ gravity in four dimensions developed in the metrif formalism. We consider perturbations by massless scalar and employ the Wentzel-Kramers-Brillouin (WKB) approximation up to sixth order to calculate the corresponding quasinormal frequencies and supplementing our findings in the eikonal approximation.

Given the simple form of the metric potentials, which yield a well-defined effective potential, we apply the standard WKB approximation directly. However, various analytical and numerical methods exist in the literature, each with its own advantages and limitations. Exact analytical solutions for black hole quasinormal mode (QNM) spectra are rare and known only for specific spacetimes, including:
i) P\"oschl-Teller potential: When the effective potential takes this exact form, QNM frequencies are given by a simple analytical expression (as detailed in later sections; see e.g., \cite{Poschl:1933zz, Ferrari:1984zz, Cardoso:2001hn, Cardoso:2003sw, Molina:2003ff, Panotopoulos:2018hua}).
ii) Hypergeometric functions: Cases where the radial wave equation transforms into Gauss's hypergeometric equation allow exact solutions (e.g., \cite{Birmingham:2001hc, Fernando:2003ai, Fernando:2008hb, Gonzalez:2010vv, Destounis:2018utr, Ovgun:2018gwt, Rincon:2018ktz}).
iii) Heun functions: For certain metrics, such as Kerr-de Sitter black holes, the equations map onto Heun's equations, enabling exact QNM determination (e.g., \cite{Hatsuda:2020sbn, Fiziev:2011mm, Naderi:2024dhh}).

Due to the complexity and non-triviality of the governing differential equations, QNM frequency calculations typically rely on numerical or semi-analytical methods.
Some examples include:
i) Frobenius Method and its generalizations \cite{Destounis:2020pjk, Fontana:2022whx, Hatsuda:2021gtn},
ii) Continued Fraction Method and its refinements \cite{Leaver:1985ax, Nollert:1993zz, Daghigh:2022uws},
iii) Asymptotic Iteration Method \cite{Cho:2011sf, 2003JPhA...3611807C, Ciftci:2005xn}.
For a more comprehensive review of these and other methods, see \cite{Konoplya:2011qq}.

\subsection{QNMs for massless scalar perturbations}

Let us consider the dynamics of a test scalar field, $\Phi$, propagating in a four-dimensional space-time background. In the following we shall assume a real scalar field. We thus consider the following action
\begin{align}
S[\Phi] \equiv \frac{1}{2} \int \mathrm{d}^4 x \sqrt{-g}
\Bigl[
\partial^{\mu} \Phi \partial_{\mu} \Phi - m^2 \Phi^2
\Bigl]\,.
\end{align}
with $m$ being the mass of the field. Based on such an action, we obtain the corresponding equation of motion for a massive scalar field \cite{Crispino:2013pya,Kanti:2014dxa,Pappas:2016ovo,Panotopoulos:2019gtn,Avalos:2023ywb,Gonzalez:2022ote,Rincon:2020cos}
\begin{equation}
\frac{1}{\sqrt{-g}}\partial_{\mu}\left(\sqrt{-g}g^{\mu\nu}\partial_{\nu}\Phi\right) =  m^2 \Phi,
\end{equation}
or, in the case of a massless scalar field the wave equation takes the form
\begin{equation}
\frac{1}{\sqrt{-g}}\partial_{\mu}\left(\sqrt{-g}g^{\mu\nu}\partial_{\nu}\Phi\right) = 0.
\end{equation}
In order to decouple the function into radial, angular and temporal contributions, let us use the symmetries of the metric, and assume an angular space spanned by spherical harmonics as eigenfunctions, to express the $\Phi$ as
\begin{equation}
\Phi(t, r, \theta, \phi) 
=\sum_{\ell ,m}e^{-i\omega t}\frac{\psi(r)}{r}Y_{\ell m}(\theta, \phi),\label{fdc}
\end{equation}
where we have also adopted the monochromatic ansatz with harmonic time dependence $e^{-i\omega t}$, corresponding to a single Fourier mode of the scalar field perturbation (with fixed multipole indices $\ell, m$). 
After applying the above ansatz, the differential equation can be written down in the following form
\begin{align}
\begin{split}
& \frac{\omega^{2}r^{2}}{A(r)} + \frac{r}{\psi(r)}\frac{\mathrm{d}}{\mathrm{d}r}\left[r^{2}A(r)\frac{\mathrm{d}}{\mathrm{d}r}\left(\frac{\psi(r)}{r}\right)\right] = \ell(\ell+1),
\label{KG}
\end{split}
\end{align}
where we have used the angular part and replace this part in terms of its eigenvalues, i.e., 
\begin{align}
    \begin{split}
&\frac{1}{\sin\theta}\frac{\partial}{\partial\theta}\left(\sin\theta\frac{\partial Y}{\partial\theta}\right) + \frac{1}{\sin^{2}\theta}\frac{\partial^{2}Y}{\partial\phi^{2}} = 
-\ell(\ell + 1)Y,
\label{kg2}
\end{split}
\end{align}
with $Y \equiv Y(\Omega) = Y(\theta, \phi)$, while  $\ell(\ell + 1)$ is the eigenvalue, and $\ell$ is the angular degree. 
After combining Eqs. (\ref{KG}) and (\ref{kg2}), we obtain a second-order differential equation for the radial coordinate. It can also be rewritten by using the so-called ``tortoise coordinate" $r_{*}$, defined as follows
\begin{equation}
\label{tcd1}
    \mathrm{d}r_{*}  \equiv \frac{\mathrm{d}r}{A(r)}\,, 
\end{equation}
Using \eqref{tcd1} into the above-mentioned second order differential equation, we obtain its canonical Schr{\"o}dinger-like form
\begin{equation} \label{SLE}
\frac{\mathrm{d}^{2}\psi(r_*)}{\mathrm{d}r_{*}^{2}} + \left[\omega^{2} - V_s(r)\right]\psi(r_*) = 0,
\end{equation}
where $V_s(r)$ is the effective potential barrier, which is computed to be
\begin{equation}
V_s(r) = A(r)
\Bigg[m^2 +
\frac{\ell(\ell + 1)}{r^{2}} + \frac{A'(r)}{r}
\Bigg],\label{poten}
\end{equation}
and where the prime represents the derivative of the radial variable with respect to the tortoise coordinate. In the case of massless scalar fields the effective potential takes the form
\begin{equation}
V_{s,m=0}(r) = A(r)
\Bigg[
\frac{\ell(\ell + 1)}{r^{2}} + \frac{A'(r)}{r}
\Bigg].
\end{equation}
Given the form of the potential, it is easy to verify that at large $r$ its asymptotic behavior is as follows
\begin{equation}
V_s(r) \sim \frac{\ell(\ell + 1)}{2 r^2} ,
\end{equation}
and so there is a numerical factor of $1/2$ compared to the Schwarzschild case, also known as the deficit angle.
Therefore the frequencies must be of the form
\begin{equation}
\omega(\ell) \propto \omega_{\text{Sch}}(\ell_{\text{eff}}) ,
\end{equation}
where the effective angular degree is computed by $\ell_{\text{eff}} (\ell_{\text{eff}}+1)=2 \ell (\ell+1)$, with $\ell$ being an integer whereas $\ell_{\text{eff}}$ being a real number. Since the spectrum of the Schwarzschild BH is known, one might think that the spectrum of the BH discussed here is also known due to the close relation to the Schwarzschild frequencies. Notice, however, that the frequencies reported in the literature are computed considering integer values of $\ell$, and therefore the numerical values of the frequencies for $\ell_{\text{eff}}$ are not known. It is for that reason that it makes sense to perform the analysis that follows.

Finally, the problem at hand is solved imposing appropriate boundary conditions. In the following we shall consider the case of a massless scalar field, $m \rightarrow 0$. In this case, since there is no cosmological constant, the effective potential barrier vanishes at infinity, $V_{s,m=0}(r) \rightarrow 0$ as $r \rightarrow \infty$, and therefore we impose as usual the following boundary conditions
\begin{align}
  \psi \rightarrow \: &\exp(-i \omega r_*), \; \; \; \; \; \;  r_* \rightarrow - \infty,
   \\
   \psi \rightarrow \: &\exp(+i \omega r_*), \; \; \; \; \; \; r_* \rightarrow + \infty .
\label{pbc}
\end{align}
for purely ingoing modes at the event horizon and purely outgoing at asymptotic infinity, respectively. 

For a perturbation with time dependence $\Phi \sim e^{-i \omega t}$, the mode decays exponentially if $\operatorname{Im}(\omega) < 0$, signifying stability against the perturbation. In contrast, $\operatorname{Im}(\omega) > 0$ produces exponential growth, indicating instability. Consequently, the sign of the imaginary part of the quasi-normal mode (QNM) frequency determines whether the black hole is stable or unstable to scalar perturbations.

\begin{figure*}[ht!]
	\centering
	\includegraphics[scale=0.618]{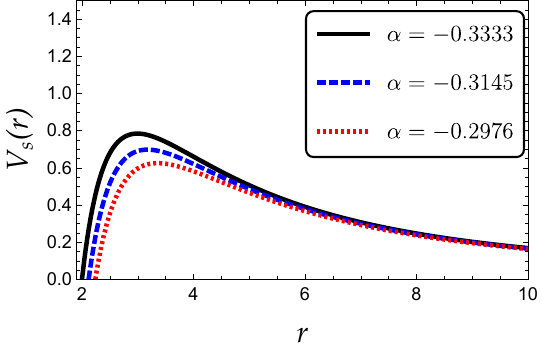} \
	\includegraphics[scale=0.618]{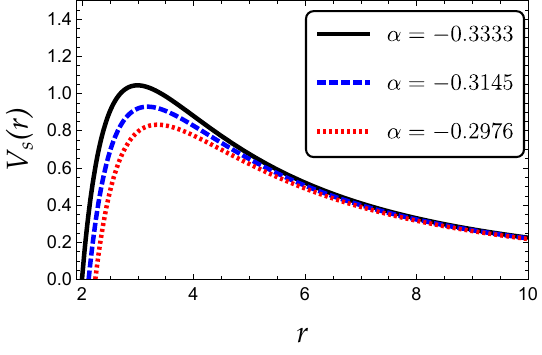} \  
    \includegraphics[scale=0.618]{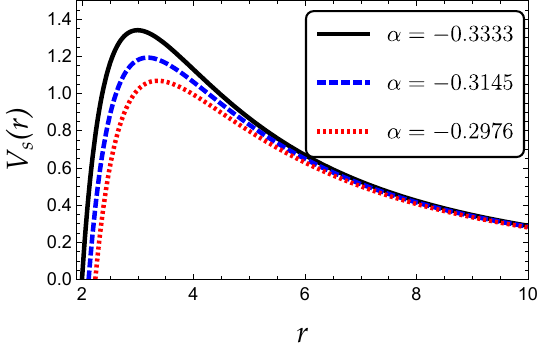} 
    \caption{
		Effective potential barrier (setting $m=0$) for different values of the parameters $\{\alpha, \ell)\}$. The value $\alpha =-1/3$ represents the usual Schwarzschild horizon (with $r_H=2$) in this representation.
        {\bf{Left panel:}} Massless scalar effective potential $V_s(r)$ against the radial coordinate $r$ for $\ell = 6$.
        {\bf{Middle panel:}} Massless scalar effective potential $V_s(r)$ against the radial coordinate $r$ for $\ell = 7$.
        {\bf{Right panel:}} Massless scalar effective potential $V_s(r)$ against the radial coordinate $r$ for $\ell = 8$.
		}
	\label{fig:1} 	
\end{figure*}

In Figure~\eqref{fig:1}, we display the effective potential barrier $V_s(r)$ for scalar perturbations as a function of the radial coordinate $r$, for various values of the angular degree, $\ell$, while keeping the free parameter $\alpha$ fixed.

From Figure~\eqref{fig:1}, we observe that as the parameter $\alpha$ becomes less negative, the maximum of the effective potential decreases while shifting towards larger radii. Additionally, all curves converge at large radii, whereas small differences persist at small radii. The same qualitative behavior is observed when the angular degree $\ell$ is increased.

Before we proceed with the analysis and the presentation of the results, a remark is in order here regarding the asymptotic structure. Contrary to the Schwarzschild geometry, the BH solution discussed here is not asymptotically flat but asymptotically conical. As far as the quantum mechanical problem is concerned, however, the details of the gravitational theory and curvature are not relevant. Only the form of the effective potential barrier matters, the asymptotic behavior of which is the same, namely $V_{s,m=0} \sim 1/r^2$, both for Schwarzschild and the solution discussed in the present work.


\subsection{The WKB approximation}

As mentioned above, the effective potential exhibits well-behaved characteristics, which justifies the use of the WKB semi-classical approximation to compute the corresponding quasi-normal mode (QNM) frequencies (see Refs.~\cite{Schutz:1985km,Iyer:1986np,Iyer:1986nq,Kokkotas:1988fm,Seidel:1989bp,Lutfuoglu:2025hjy,Hamil:2024njs,Hamil:2024ppj} and references therein).
The method was originally introduced by Schutz and Will at first order~\cite{Schutz:1985km} and later extended by Iyer and Will to include second- and third-order corrections~\cite{Iyer:1986np}. Significant advances were subsequently made by R.~A.~Konoplya, who developed the approximation up to sixth order~\cite{Konoplya:2003ii}, and by Matyjasek and Opala, who further extended it to thirteenth order~\cite{Matyjasek:2017psv}, thereby greatly improving its accuracy and applicability to a broader range of potentials.
Most conveniently used at lower-order formulations, it has proven highly effective for computing the fundamental quasi-normal mode (QNM) of black holes, especially in the Schwarzschild case. Its accuracy increases with increasing values of the angular harmonic index $\ell$, while it tends to decrease for higher overtone numbers $n$.

It should be noted that the convergence of higher-order WKB approximations lacks a rigorous mathematical proof. As a heuristic, Konoplya suggested estimating errors by comparing QNM frequencies across successive WKB orders, a practical but non-rigorous method for selecting the optimal order. By experience, the sixth or seventh-order approximations typically yield the most accurate results, though this depends strongly on the background geometry, precluding a universally optimal order.

The well-known technique begins by solving a one-dimensional Schr{\"o}dinger-like equation with an effective potential. It involves matching asymptotic solutions through a Taylor expansion around the peak of the potential at $x = x_0$; the point where the standard WKB approximation breaks down. This expansion remains valid in the region between the two turning points, which are defined as the roots of the equation $U(x, \omega) \equiv V(x) - \omega^2 = 0$.

In this paper, we apply the WKB method to compute QNM  frequencies up to the sixth order, using the generalized formula:
\begin{equation}
\omega_n^2 = V_0 + \sqrt{-2V_0''}\, \Lambda(n) - i\, \nu \sqrt{-2V_0''} \left[1 + \Omega(n)\right], 
\end{equation} 
where the set $\{V_0, V_{0}'', \nu, \Lambda(n), \Omega(n)\}$ have the usual meaning, i.e.:
(a) $V_0$ being the maximum of the effective potential, 
(b) $V_0''$ represent its second derivative at the maximum, 
(c) $\nu = n + 1/2$ with $n=0,1,2,\dots$ the overtone number, and finally
(d) $\Lambda(n)$ and $\Omega(n)$ are correction functions explicitly given in \cite{Kokkotas:1988fm}.  
We utilize the Wolfram Mathematica implementation of the WKB method \cite{wolfram}, which natively supports orders from one to six \cite{Konoplya:2019hlu} and has been extended up to the thirteenth order. In our analysis, we restrict ourselves to the regime where $n < \ell$ in order to ensure the validity and reliability of the approximation. For higher-order corrections, we refer the reader to Refs.~\cite{Konoplya:2019hlu,Hatsuda:2019eoj}.


\begin{figure*}[t!]
	\centering
	\includegraphics[scale=0.934]{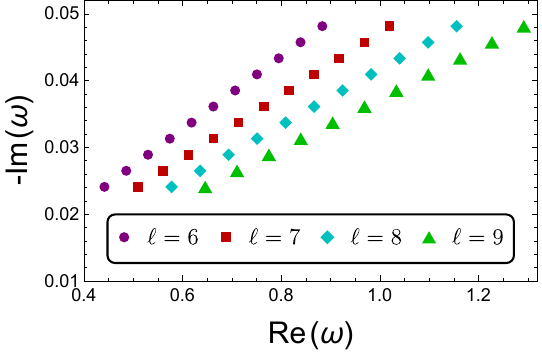} \
	\includegraphics[scale=0.934]{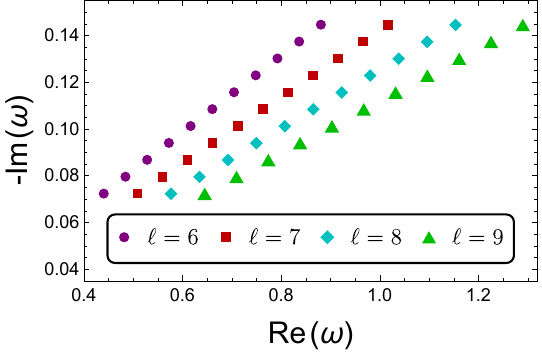} \  
    \caption{
	Quasi-normal modes (QNMs) for massless scalar perturbations. 
    {\bf{Left panel:}} QNMs corresponding to running values of $\alpha$ and different values of the angular momentum quantum number $\ell$, for the fundamental mode ($n=0$).
     {\bf{Right panel:}} QNMs corresponding to running values of $\alpha$ and different values of the angular momentum quantum number $\ell$, for the first excited mode $(n=1)$.
	}
	\label{fig:2} 	
\end{figure*}


\begin{figure*}[t!]
	\centering
	\includegraphics[scale=0.934]{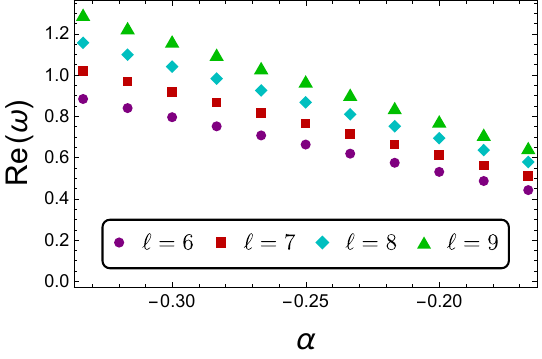} \
	\includegraphics[scale=0.934]{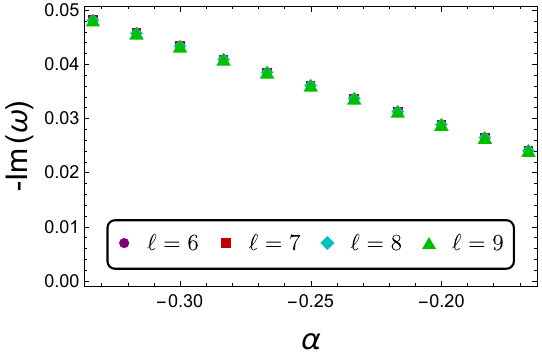} \  
    \\
        \includegraphics[scale=0.934]{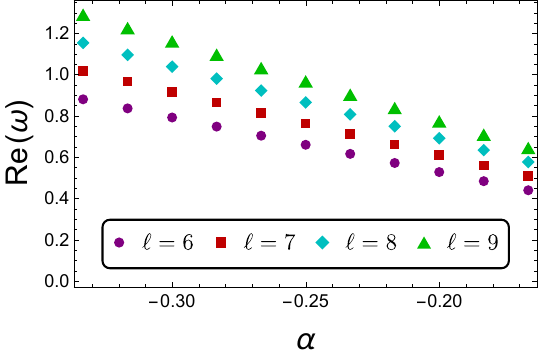} \
	\includegraphics[scale=0.934]{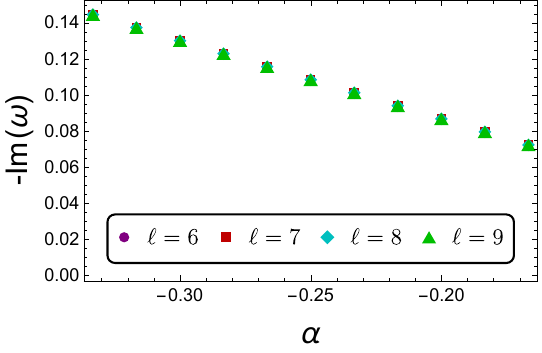} \  
    \caption{Real/imaginary part of scalar QNM frequencies, $\omega_R$/$\omega_{I}$, against the parameter $\alpha$.
		{\bf{Top Left Panel:}} $\text{Re}(\omega)$ vs $\alpha$, varying $\ell$ from 6 to 9  for the fundamental mode $(n=0)$.
		{\bf{Top Right Panel:}} $-\text{Im}(\omega)$ vs $\alpha$, varying $\ell$ from 6 to 9  for the fundamental mode $(n=0)$.
        {\bf{Down Left Panel:}} $\text{Re}(\omega)$ vs $\alpha$, varying $\ell$ from 6 to 9 for the first exited mode $(n=1)$.
		{\bf{Down Right Panel:}} $-\text{Im}(\omega)$ vs $\alpha$, varying $\ell$ from 6 to 9 for the first excited mode $(n=1)$.
	}
	\label{fig:3} 	
\end{figure*}


\subsection{On analytic expressions for QNMs}

Following the spirit of the WKB approximation, which is known to converge only asymptotically and to provide exact results solely in the eikonal (large angular momentum) limit, Konoplya and Zhidenko recently demonstrated \cite{Konoplya:2023moy} a non-trivial extension that allows one to derive accurate analytical expressions for the quasi-normal modes and grey-body factors of black holes by systematically going beyond the eikonal approximation.
After applying this underlying idea to our specific case, the procedure complements our numerical results and yields excellent agreement with them.
This idea has been applied satisfactorily in some papers,  see for instance \cite{Malik:2024voy,Malik:2024sxv,Koch:2025gaw} and references therein.

For simplicity, let us focus on the underlying idea and the necessary expressions. 
The effective potential can be approximated by the expression:
\begin{equation}\label{potential-multipole}
V(r_*)=\kappa^2\Bigl(H(r_*)+\Order{\kappa^{-1}}\Bigl),
\end{equation}
where $\kappa\equiv\ell+1/2$. 
Having the effective potential, we perform an expansion of the position of its maximum and the frequency in terms of $\kappa^{-1}$.
Given that the function $H(r_*)$ has a maximum only, its position (see Eq.(\ref{potential-multipole})) can be written as follows:
\begin{equation}\label{rmax}
  r_{\max }=r_0+r_1\kappa^{-1}+r_2\kappa^{-2} + \cdots .
\end{equation}
As an elementary check, let us use (\ref{rmax}) into the following first order WKB formula for the frequency
\begin{eqnarray}
\omega&=&\sqrt{V_0-\imo \K\sqrt{-2V_2}},
\end{eqnarray}
and after that expanding it in terms of powers of $\kappa^{-1}$, it is possible to obtain \cite{Konoplya:2023moy}
\begin{eqnarray}\label{eikonal-formulas}
\omega=\Omega\kappa-\imo\lambda\K+\Order{\kappa^{-1}}.
\end{eqnarray}
It becomes at this point obvious that: 
 i) $\Omega$ symbolize the angular velocity at the unstable null geodesics, and 
ii) $\lambda$ means the Lyapunov exponent. 
Notice that we have used the definition $\K \equiv \nu = n+1/2$.
At this point, we cannot overlook the opportunity to compare our derived expression with the now-famous formulas presented in Eq.~\eqref{eikonal-formulas}, which were first obtained by V.~Cardoso and collaborators \cite{Cardoso:2008bp}. In that work, the authors established a connection between the unstable null geodesics surrounding a static, spherically symmetric, asymptotically flat or de Sitter black hole and its quasi-normal mode frequencies in the regime where $\ell \gg n$.
The expression is written via the equation \cite{Cardoso:2008bp}:
\begin{equation}\label{QNM}
\omega_n=\Omega\ell-\imo\bigg(n + \frac{1}{2}\bigg)|\lambda|, \quad \ell \rightarrow \infty.
\end{equation}
It should be noted that, as pointed out in Ref.~\cite{Konoplya:2017wot}, there exist cases in which this correspondence is violated. Particularly notable examples include the Einstein-(dilaton)-Gauss-Bonnet theory and Einstein-Lovelock gravity \cite{Konoplya:2017wot,Konoplya:2020bxa,Konoplya:2019hml}.
The analytic expression for the scalar case, considering up to fourth order beyond eikonal limit is written as follows:
\begin{widetext}
\begin{align} \label{wanaliticscalar}
\begin{split}
\omega = 
-\frac{\alpha  \kappa }{\sqrt{6}}+\frac{i \alpha  \K}{2 \sqrt{3}}+\frac{5 \alpha  \left(12 \K^2+5\right)}{864 \sqrt{6} \kappa }+\frac{i \alpha  \K \left(8716 \K^2+313\right)}{62208 \sqrt{3} \kappa ^2}+\frac{\alpha  \left(76176 \K^4+52968 \K^2+6577\right)}{1492992 \sqrt{6} \kappa ^3} + \mathcal{O}\left(\left(\frac{1}{\kappa }\right)^5\right) ,
\end{split}
\end{align}
where we have performed an expansion in powers of $  \kappa^{-1}  $, following the approach introduced in Ref.~\cite{Konoplya:2023moy}. In this way, we obtain the series expansion for the location of the effective potential maximum in the case of a massless scalar field, given by:
\begin{align}\label{rmax-scalar}
r_{\max } &= -\frac{1}{\alpha }+\frac{1}{18 \alpha  \kappa ^2}-\frac{7}{648 \alpha  \kappa ^4} + \mathcal{O} \left(\left(\frac{1}{\kappa }\right)^5\right).
\end{align}
\end{widetext}
Table~\eqref{tab:3} contains the frequencies of the fundamental mode using the analytical expression. After comparing the numerical results obtained by using the WKB semi-analytical method to 6th order and the above mentioned analytical expression, we confirm an excellent agreement between the results for $\ell > 0$ (see also the purple highlighted rows in Table~\ref{tab:1} in the appendix for a detailed comparison). 
To be more precise, our results demonstrate that both the real and imaginary parts of the quasi-normal mode frequency exhibit nearly perfect agreement with the WKB approximation as the parameter $\alpha$ increases. Consistent with the WKB predictions, the real part $\operatorname{Re}(\omega)$ decreases with increasing $  \alpha  $, while the imaginary part $-\operatorname{Im}(\omega)$, which corresponds to the decay rate, also decreases.

Tables \ref{tab:5} show the percentage error between the WKB method and the analytic expressions beyond the eikonal approximation.
We have used the usual expression of percent error, i.e.,  
\begin{align} \label{ErrorR}
    \delta_R(\omega_R) &\equiv \Bigg| \frac{\omega_R^{\text{WKB}} - \ \omega_R^{\text{analytic}}}{\omega_R^{\text{analytic}}}\Bigg|\times100\% ,
    \\
    \label{ErrorI}
      \delta_R(\omega_I) &\equiv \Bigg| \frac{ \omega_I^{\text{WKB}} - \ \omega_I^{\text{analytic}} }{ \omega_I^{\text{analytic}} }\Bigg|\times100\%,
\end{align}
where we also define the error, component by component, using
\begin{align}\label{Error}
    \delta_R &\equiv (\delta_R(\omega_R) , \delta_R(\omega_I)).
\end{align}
From the tables, we can confirm that the error in the real part of the quasinormal frequencies reaches a maximum of approximately $0.00096\%$, occurring for large values of $\alpha$ and small values of $\ell$, as expected. 
Thus, the real part is essentially identical between the two methods, meaning that the oscillation frequency of the perturbation is insensitive to the choice of approach. 
Additionally, the error in the imaginary part reaches a maximum of $\sim 0.14\%$, again for large $\alpha$ and small $\ell$. In both the real and imaginary parts, the error is reduced when $\ell$ increases, consistent with the improved performance of the WKB approximation at higher multipoles.
In summary, we have confirmed numerically that both the WKB and analytic methods yield highly accurate results, at least for the parameter values considered in this work.

\subsection{The eikonal limit}

In the context of black hole physics, the eikonal approximation (i.e., the geometrical optics approximation) is a powerful high-frequency/semi-classical tool widely used for studying quasi-normal modes.
The eikonal limit corresponds to perturbations with very large multipole number $\ell \gg 1$ (high angular momentum quantum number). Under such circumstance:
(i) the wavelength of the perturbation is much shorter than the typical curvature scale of the spacetime (e.g., the black hole's horizon or photon sphere radius), and
(ii) wave propagation behaves like rays in geometrical optics, analogous to light rays in optics or high-frequency limits in quantum mechanics (WKB approximation).
Thus, the problem becomes tractable analytically, as the full wave equation reduces to simpler transport equations along null geodesics.

For a static, spherically symmetric black hole, the complex QNM frequency in the eikonal limit is found to be:
\begin{align}
\omega_{\text{QNM}} \approx \Omega_c \ell - i \left(n + \frac{1}{2}\right) |\lambda|,
\end{align}
where the the symbols involved have the usual meaning, i.e., 
(i) $\Omega_c$ represent the angular frequency of the unstable circular photon orbit at radius $r_c$ (the so-called "photon sphere"),
(ii) $\lambda$ is the Lyapunov exponent, measuring the instability timescale of that orbit (representing how quickly nearby geodesics diverge),
(iii) $n = 0,1,2,\dots$ is the overtone number (being $n=0$ the fundamental mode), 
(iv) the real part $\operatorname{Re}(\omega) \approx \ell \Omega_c$ (formarly $\operatorname{Re}(\omega) = \Omega_c \kappa$) gives the oscillation frequency and
(v) the imaginary part $  \operatorname{Im}(\omega) \approx -\left(n + \frac{1}{2}\right) |\lambda|  $ gives the decay rate (negative for stability).

This approach links gravitational-wave ringdown physics directly to classical geodesic motion making this formulation a beautiful geometric interpretation.

Focusing on the concrete computation, when $ \ell \rightarrow \infty$, the angular momentum term dominates in the expression for the effective potential and the formula is reduce to
\begin{equation}
V(r) \approx \frac{A(r) \ell^2}{r^2} \equiv \ell^2 g(r) ,
\end{equation}
where for simplicity we have defined a new function $g(r) \equiv A(r)/r^2$.
The maximum of the potential $r_1$ can be obtained after solving the following algebraic equation
\begin{equation}
2 A(r_1) - r_1 A'(r)|_{r_1} = 0.
\end{equation}
We have completed our study by computing (i) $\Omega_c \propto \omega_R$ and (ii) $|\lambda| \propto \omega_I$ by using the following analytic expressions: for the Lyapunov exponent $\lambda$ is given by \cite{Cardoso:2008bp}
\begin{equation}
 \lambda  = r_1^2 \sqrt{\frac{g''(r_1) g(r_1)}{2}} = - i\frac{\alpha }{2 \sqrt{3}},
\end{equation}
while for the angular velocity $\Omega_c$ at the unstable null geodesic is given by \cite{Cardoso:2008bp}
\begin{equation}
\Omega_c = \frac{\sqrt{f(r_1)}}{r_1} = -\frac{\alpha }{\sqrt{6}}.
\end{equation}
At this point it is worth mentioning that it is not difficult to see that the first derivative of the gravitational Lagrangian, $f_R=1+f'(R)=1+\alpha r$, vanishes precisely at the photon orbit, $r_{ph}=-1/\alpha > 0$, which is larger than the horizon. This indicates that the slogan "The QN frequencies are determined by the photon sphere" may fail. Despite that, one can still claim that the same expressions for $\{ \Omega_c, \lambda \}$ are obtained within the WKB approximation of 1st order, see e.g. \cite{Ponglertsakul:2018smo} for details. Moreover, since $f_R$ controls the effective 
gravitational coupling and the scalar degree of freedom in the Einstein frame, the vanishing of $f_R$ may directly affect gravitational stability, although this must be carefully studied via perturbation theory, which lies outside the scope of this work.

\begin{figure*}[ht!]
\centering
\includegraphics[scale=0.96]{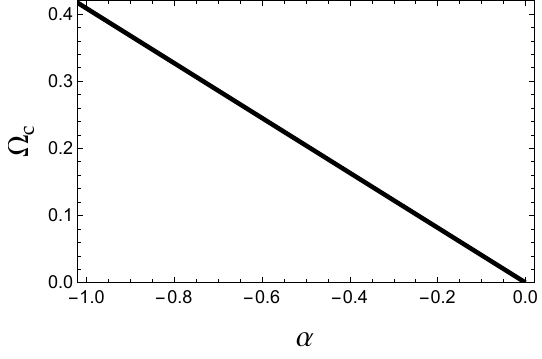} \
\includegraphics[scale=0.96]{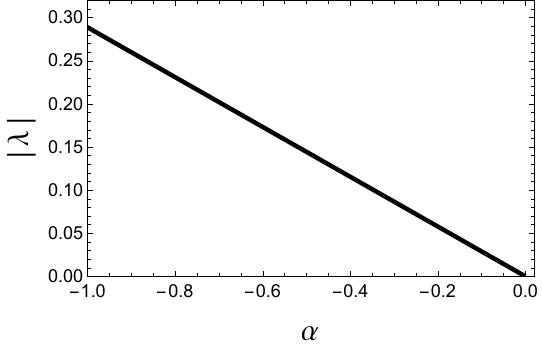} \
\caption{
QNMs in the eikonal limit:
{\bf{Left panel:}} Angular velocity vs the parameter $\alpha$.
{\bf{Right panel:}} Lyapunov exponent against the parameter $\alpha$.
}
\label{fig:eikonal} 	
\end{figure*}

In Figure~\ref{fig:eikonal}, we display (i) the angular frequency $  \Omega_c  $ (left panel) and (ii) the absolute value of the Lyapunov exponent $ | \lambda | $ (right panel) as functions of the parameter $  \alpha  $. Our results show excellent agreement with the two independent computations presented earlier.


\section{Absorption cross section}
\label{S:IV}

The GBF of the BH is given by $|T(\omega)|^2$ which is the probability of the wave tunneling through the potential created by the field in consideration. Here, we employ the WKB approximation developed by Schutz and Will \cite{Schutz:1985km}, and Iyer and Will \cite{Iyer:1986np} to compute the GBFs and partial absorption cross sections. 
The WKB approximation has high accuracy when $\omega^2 \approx V_0$ where 
(i) $V_0$ is the potential at $r=r_0$ and 
(ii) $r_0$ is the value when the effective potential take the maximum. In the WKB approximation, the reflection coefficient $R(\omega)$ is given by
\begin{equation} 
R(\omega) = (1+e^{-2\pi i b})^{-1/2},
\, 
\label{Refl}
\end{equation}
and 
\begin{equation}
|T(\omega)|^2 = 1 - |R(\omega)|^2 ,
\, 
\label{}
\end{equation}
where $T(\omega)$ is the transmission coefficient. The value $b$ is eq.$\eqref{Refl}$, which accounts the order of the WKB approximation, is given by 
\begin{equation}
b = i \frac{\omega^2 - V_0}{\sqrt{- 2 V''_0}} -\Lambda_2 - \Lambda_3 - \Lambda_4 - \Lambda_5 - \Lambda_6
\, .
\label{}
\end{equation}
The terms $  \Lambda_2  $, $  \Lambda_3  $, $  \Lambda_4  $, $  \Lambda_5  $, and $  \Lambda_6  $ appearing in the preceding equation are intricate expressions; their detailed definitions are provided in the seminal works on higher-order WKB approximations. 

The flux spectrum, the differential energy emission rate and the total absorption cross section are given by the following expressions, respectively \cite{Crispino:2013pya}
\begin{eqnarray}
\frac{\mathrm{d}N}{\mathrm{d}t} & = & \frac{d \omega}{2 \pi} \frac{1}{e^{\omega/T_H} - 1} \sum_{\ell=0}^\infty (2 \ell+1) \: |T_{\ell}(\omega)|^2 \\
\frac{\mathrm{d^2}E}{\mathrm{d}t \mathrm{d} \omega} & = & \frac{1}{2 \pi} \frac{\omega}{e^{\omega/T_H} - 1} \sum_{\ell} (2 \ell+1) \: |T_{\ell}(\omega)|^2 \\
\sigma & = & \sum_{\ell = 0}^\infty \frac{\pi}{\omega^2} (2 \ell+1) \: |T_{\ell}(\omega)|^2 = \sum_{\ell = 0}^\infty \sigma_{\ell}(\omega)
\end{eqnarray}
where the partial absorption cross section, $\sigma_\ell(\omega)$ is computed by
\begin{equation}
\sigma_{\ell} = \frac{\pi}{\omega^2} (2 \ell + 1)  |T_{\ell}(\omega)|^2
\, .
\label{}
\end{equation}
while the Hawking temperature is computed to be $T_H=3(-\alpha)/(16 \pi )$.

The WKB approximation for calculating the reflection coefficient $  R(\omega)  $, transmission coefficient $  T(\omega)  $, and greybody factors has been widely applied in the literature, including in Refs.~\cite{Toshmatov:2016bsb,Koch:2025gaw}. Consequently, we do not repeat the basic computational details here, as the expressions given above are adequate for our purposes.


%
\begin{figure*}[ht!]
\centering
\includegraphics[scale=0.95]{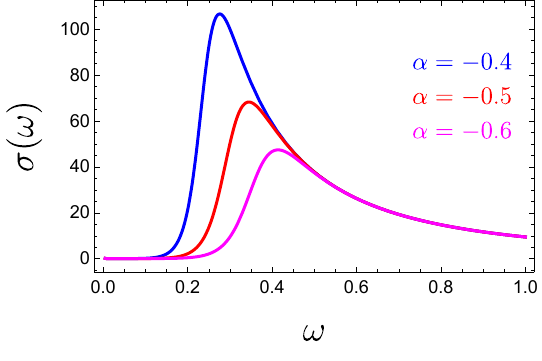} \
\includegraphics[scale=0.97]{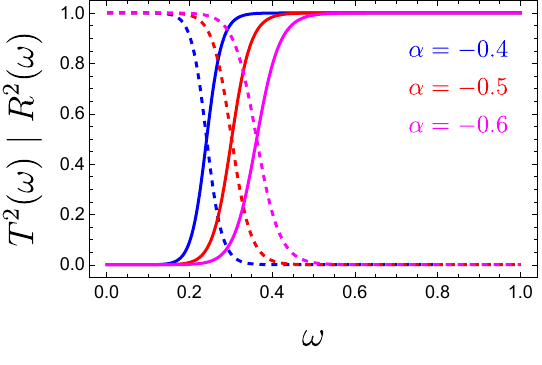} \
\\
\includegraphics[scale=0.94]{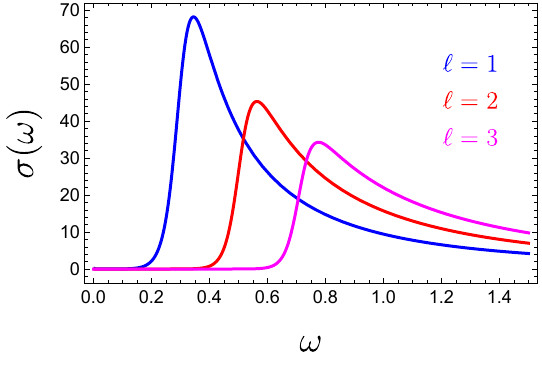} \
\includegraphics[scale=0.96]{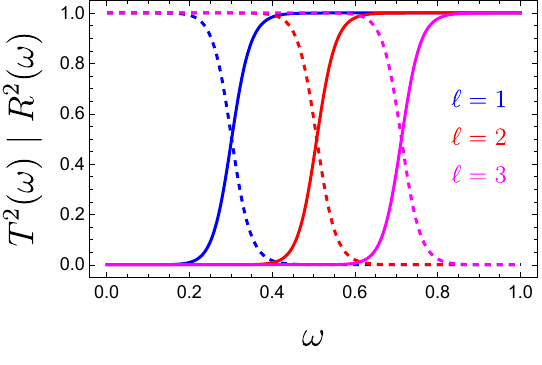} \
\caption{
The figure displays the partial absorption cross section $\sigma_\ell(\omega)$ (left panel) and the coefficients $|R|^2,|T|^2$ (right panel) for different values of the parameters $\{\alpha, \ell \}$. All plots correspond to the case of a massless scalar field.
For the first row we consider the values $m = 0$, $\ell=1$ and three different values of the parameter $\alpha$.
For the second row we consider the values $m = 0$, $\alpha=-0.5$ and three different values of the angular momentum $\ell$.
}
\label{fig:GBF_vary_parameters} 	
\end{figure*}

\section{Discussion of results}
\label{S:V}

We summarize our main numerical results in 3 tables and 5 figures. 
First, Figure~\eqref{fig:1} displays the effective potential for massless scalar perturbations considering three different values of the angular degree, $\ell=6,7,8$, as well as three different values of the parameter $\alpha=-0.3333, -0.3145, -0.2976$. The potential has the typical shape of a potential barrier, it vanishes at the horizon and at infinity, while at the same time it reaches its global maximum at some value of the radial coordinate in the range $[r_H, \infty)$. Increasing $\ell$ shifts the curve upwards, whereas increasing $\alpha$ shifts the curve downwards.

Next, in Figures \eqref{fig:2} and \eqref{fig:3} we show the frequencies of the QNMs for massless scalar perturbations considering four different values of the angular degree, $\ell=6,7,8,9$, two different values of the overtone, namely $n=0$ (fundamental mode) and $n=1$ (first excited mode) and eleven different values of the parameter $\alpha$, from $-0.3333$ up to $-0.1666$, shown in Table I. All frequencies computed here are found to be complex numbers with a positive real part and a negative imaginary part, and therefore all the modes computed in this work are stable. 
Figure~\eqref{fig:2} shows the negative value of the imaginary part versus the real part, while Figure~\eqref{fig:3} displays both the real part and the negative of imaginary part versus $\alpha$. The real part increases with $\ell$, decreases with $\alpha$, and is insensitive to $n$. Regarding the absolute value of the imaginary part, it decreases with $\alpha$, increases with $n$, and depends only slightly on $\ell$ (almost insensitive to it).

The case of the eikonal limit, $\ell \gg 1$, is considered separately in Figure~\eqref{fig:eikonal}, in which we show the critical frequency, $\Omega_c$, which determines the real part of the frequencies in the limit $\ell \rightarrow \infty$, as well as the Lyapunov exponent, $\lambda$, which determines the imaginary part of the frequencies in the limit $\ell \rightarrow \infty$, as a function of the parameter $\alpha$. Here, too, it is observed one more time that both $\Omega_c$ and $\lambda$ decrease with $\alpha$.

Finally, the results on the GBFs for massless scalar perturbations are shown in Figure~\eqref{fig:GBF_vary_parameters}. In that, we display the absorption cross section as well as the square of transmission and reflection coefficients, $\{ T(\omega), R(\omega)\}$, as functions of the frequency assuming three different values of the angular degree, $\ell$, 
and the parameter $\alpha$. 
To be more precise, we have considered the cases $\ell=1,2,3$ 
and $\alpha=-0.6,-0.5,-0.4$. 
It is observed that with increasing $\ell$ and $|\alpha|$ the maximum of the absorption cross section decreases, while at the same time it is located at a higher frequency. 
%
Regarding the reflection and transmission coefficients, they take values in the range from zero to unity, with $|T|^2$ being a monotonically increasing function of $\omega$, whereas $|R|^2$ is a monotonically decreasing function of the frequency. It is observed that increasing $\ell$ and $|\alpha|$ shifts the curves to the right.

\section{Summary and concluding remarks}
\label{S:VI}

To summarize, we have investigated several nontrivial and physically significant features from a purely gravitational perspective within the framework of a specific four-dimensional $f(R)$ gravity model. After a concise introductory section outlining the background spacetime and the theoretical model under consideration, we presented the essential theoretical framework required to contextualize and interpret our results.

We began by reviewing the fundamental theory of QNMs, including the WKB approximation, the eikonal limit, and the corresponding analytical expressions. These three complementary approaches consistently validated our results and demonstrated that the black hole solution is stable under massless scalar perturbations.

Subsequently, we computed closely related observables derived from the WKB analysis, namely the absorption cross section, $\sigma_\ell(\omega)$, and the reflection and transmission coefficients, $|R|^2, |T|^2$. The absorption cross section was presented graphically for various values of the relevant parameters. In general, the results show that: (i) as the parameter $\alpha$ decreases towards more negative values, $\sigma(\omega)$ decreases and its maximum shifts toward higher frequencies;
(ii) increasing the multipole number $\ell$ leads to a decrease in $\sigma_\ell(\omega)$, with a more pronounced shift of the maximum toward higher frequencies, mirroring the behavior observed for negative $\alpha$. 

Finally, as far as the reflection and transmission coefficients are concerned, they take values in the range from zero to unity, with $|T|^2$ being a monotonically increasing function of $\omega$, whereas $|R|^2$ is a monotonically decreasing function of the frequency. Our results show that increasing $\ell$ and $|\alpha|$ shifts the curves to the right.
%

\vskip 0.33cm


\vskip 0.33cm


\section*{Acknowledgments}

We wish to thank the anonymous reviewer for useful comments and suggestions. A.~R. would like to express his gratitude to Silesian University in Opava, Czech Republic, for their financial support.
A.~R. is very grateful for the hospitality of the University of Valencia (Spain), Valencia Polytechnic University (Spain) and
the Complutense University of Madrid (Spain). 
The creation of this article was supported by the grant program Vouchers for Universities in the Moravian-Silesian Region (registration number CZ.10.03.01/00/23\_042/00003901119).
This article is based upon work from COST Action FuSe, CA24101, supported by COST (European Cooperation in Science and Technology).

\section*{Data Availability Statement}
\noindent
\verb|The data generated in this paper is presented|
\verb|in tables and figures within the main body of| 
\verb|the text.|

\vskip 1cm

\appendix

\section{Computation of quasi-normal modes} \label{A:QNM}

In this Appendix, we present the computed quasi-normal modes (QNMs) for massless scalar perturbations in Table~\eqref{tab:1}. We consider values of the parameter $  \alpha  $ in the interval $\alpha \in [-0.333333, -0.166667]$. For each overtone number $n = 0,1,2,3$, we tabulate the real and imaginary parts of the QNM frequencies corresponding to angular momentum quantum numbers $\ell = 6,7,8,9,10$ in the scalar case.
Additionally, in Table~\eqref{tab:3}, we provide the QNM frequencies obtained using the exact analytic expressions for the scalar case.
In the eikonal approximation, we restrict the parameter range to $\alpha \in [-1, 0]$. Given the simplicity of the analytic form in this limit, we present only a figure to demonstrate that the behavior is in perfect agreement with our previous numerical results.

\begin{table*}
    \centering
	\caption{Massless scalar quasinormal modes using the sixth-order WKB approximation.}
	\label{tab:1}       
	\begin{tabular}{c|c|c|c|c|c|c}
		\hline\noalign{\smallskip}
		$\alpha$  & $n$ & $\ell=6$ & $\ell=7$ & $\ell=8$ & $\ell=9$ & $\ell=10$  \\
		\noalign{\smallskip}\hline\noalign{\smallskip}
          \noalign{\smallskip}\hline
            \rowcolor{purple!30}
 -0.333333 & 0 & 0.883569\, -0.0481326 i & 1.019780\, -0.0481276 i & 1.155960\, -0.0481243 i & 1.292120\, -0.0481219 i & 1.428270\, -0.0481202 i \\
 -0.333333 & 1 & 0.880672\, -0.1446040 i & 1.017270\, -0.1445380 i & 1.153740\, -0.1444940 i & 1.290140\, -0.1444620 i & 1.426470\, -0.1444400 i \\
 -0.333333 & 2 & 0.874944\, -0.2416920 i & 1.012280\, -0.2414120 i & 1.149340\, -0.2412240 i & 1.286190\, -0.2410920 i & 1.422890\, -0.2409960 i \\
 -0.333333 & 3 & 0.866512\, -0.3397940 i & 1.004910\, -0.3390510 i & 1.142790\, -0.3385520 i & 1.280310\, -0.3382020 i & 1.417560\, -0.3379460 i \\
		\noalign{\smallskip}\hline
          \rowcolor{purple!30}
 -0.316667 & 0 & 0.839390\, -0.0457260 i & 0.968792\, -0.0457212 i & 1.098160\, -0.0457181 i & 1.227520\, -0.0457158 i & 1.356860\, -0.0457142 i \\
 -0.316667 & 1 & 0.836639\, -0.1373740 i & 0.966405\, -0.1373110 i & 1.096060\, -0.1372690 i & 1.225630\, -0.1372390 i & 1.355150\, -0.1372180 i \\
 -0.316667 & 2 & 0.831197\, -0.2296070 i & 0.961670\, -0.2293410 i & 1.091870\, -0.2291630 i & 1.221880\, -0.2290380 i & 1.351750\, -0.2289460 i \\
 -0.316667 & 3 & 0.823186\, -0.3228040 i & 0.954668\, -0.3220980 i & 1.085650\, -0.3216250 i & 1.216290\, -0.3212920 i & 1.346680\, -0.3210490 i \\
		\noalign{\smallskip}\hline
          \rowcolor{purple!30}
 -0.300000 & 0 & 0.795212\, -0.0433194 i & 0.917803\, -0.0433149 i & 1.040370\, -0.0433118 i & 1.162910\, -0.0433097 i & 1.285440\, -0.0433082 i \\
 -0.300000 & 1 & 0.792605\, -0.1301440 i & 0.915541\, -0.1300840 i & 1.038370\, -0.1300440 i & 1.161120\, -0.1300160 i & 1.283820\, -0.1299960 i \\
 -0.300000 & 2 & 0.787450\, -0.2175230 i & 0.911056\, -0.2172700 i & 1.034400\, -0.2171020 i & 1.157570\, -0.2169830 i & 1.280600\, -0.2168970 i \\
 -0.300000 & 3 & 0.779861\, -0.3058140 i & 0.904422\, -0.3051460 i & 1.028510\, -0.3046970 i & 1.152280\, -0.3043810 i & 1.275800\, -0.3041510 i \\
		\noalign{\smallskip}\hline
          \rowcolor{purple!30}
 -0.283333 & 0 & 0.751033\, -0.0409127 i & 0.866813\, -0.0409085 i & 0.982568\, -0.0409056 i & 1.098300\, -0.0409036 i & 1.214030\, -0.0409022 i \\
 -0.283333 & 1 & 0.748572\, -0.1229140 i & 0.864678\, -0.1228570 i & 0.980682\, -0.1228200 i & 1.096620\, -0.1227930 i & 1.212500\, -0.1227740 i \\
 -0.283333 & 2 & 0.743702\, -0.2054380 i & 0.860442\, -0.2052000 i & 0.976935\, -0.2050400 i & 1.093260\, -0.2049280 i & 1.209460\, -0.2048470 i \\
 -0.283333 & 3 & 0.736535\, -0.2888250 i & 0.854176\, -0.2881930 i & 0.971375\, -0.2877690 i & 1.088260\, -0.2874710 i & 1.204930\, -0.2872540 i \\
		\noalign{\smallskip}\hline
          \rowcolor{purple!30}
 -0.266667 & 0 & 0.706855\, -0.0385061 i & 0.815824\, -0.0385021 i & 0.924770\, -0.0384994 i & 1.033700\, -0.0384975 i & 1.142620\, -0.0384962 i \\
 -0.266667 & 1 & 0.704538\, -0.1156830 i & 0.813814\, -0.1156300 i & 0.922995\, -0.1155950 i & 1.032110\, -0.1155700 i & 1.141180\, -0.1155520 i \\
 -0.266667 & 2 & 0.699955\, -0.1933540 i & 0.809828\, -0.1931290 i & 0.919468\, -0.1929790 i & 1.028950\, -0.1928740 i & 1.138310\, -0.1927970 i \\
 -0.266667 & 3 & 0.693209\, -0.2718350 i & 0.803931\, -0.2712410 i & 0.914235\, -0.2708420 i & 1.024250\, -0.2705610 i & 1.134050\, -0.2703570 i \\
		\noalign{\smallskip}\hline
          \rowcolor{purple!30}
 -0.250000 & 0 & 0.662676\, -0.0360995 i & 0.764835\, -0.0360957 i & 0.866972\, -0.0360932 i & 0.969092\, -0.0360914 i & 1.071200\, -0.0360902 i \\
 -0.250000 & 1 & 0.660504\, -0.1084530 i & 0.762951\, -0.1084030 i & 0.865308\, -0.1083700 i & 0.967603\, -0.1083470 i & 1.069850\, -0.1083300 i \\
 -0.250000 & 2 & 0.656208\, -0.1812690 i & 0.759213\, -0.1810590 i & 0.862002\, -0.1809180 i & 0.964639\, -0.1808190 i & 1.067170\, -0.1807470 i \\
 -0.250000 & 3 & 0.649884\, -0.2548450 i & 0.753685\, -0.2542880 i & 0.857096\, -0.2539140 i & 0.960232\, -0.2536510 i & 1.063170\, -0.2534590 i \\
		\noalign{\smallskip}\hline
          \rowcolor{purple!30}
 -0.233333 & 0 & 0.618498\, -0.0336928 i & 0.713846\, -0.0336893 i & 0.809174\, -0.0336870 i & 0.904486\, -0.0336853 i & 0.999788\, -0.0336842 i \\
 -0.233333 & 1 & 0.616471\, -0.1012230 i & 0.712088\, -0.1011770 i & 0.807621\, -0.1011450 i & 0.903096\, -0.1011240 i & 0.998530\, -0.1011080 i \\
 -0.233333 & 2 & 0.612461\, -0.1691840 i & 0.708599\, -0.1689880 i & 0.804535\, -0.1688570 i & 0.900330\, -0.1687650 i & 0.996024\, -0.1686970 i \\
 -0.233333 & 3 & 0.606558\, -0.2378560 i & 0.703439\, -0.2373360 i & 0.799956\, -0.2369860 i & 0.896217\, -0.2367410 i & 0.992292\, -0.2365620 i \\
		\noalign{\smallskip}\hline
          \rowcolor{purple!30}
 -0.216667 & 0 & 0.574320\, -0.0312862 i & 0.662857\, -0.0312829 i & 0.751375\, -0.0312808 i & 0.839880\, -0.0312793 i & 0.928375\, -0.0312781 i \\
 -0.216667 & 1 & 0.572437\, -0.0939928 i & 0.661224\, -0.0939497 i & 0.749933\, -0.0939208 i & 0.838589\, -0.0939006 i & 0.927207\, -0.0938859 i \\
 -0.216667 & 2 & 0.568714\, -0.1571000 i & 0.657985\, -0.1569180 i & 0.747068\, -0.1567960 i & 0.836021\, -0.1567100 i & 0.924880\, -0.1566480 i \\
 -0.216667 & 3 & 0.563233\, -0.2208660 i & 0.653194\, -0.2203830 i & 0.742816\, -0.2200590 i & 0.832201\, -0.2198310 i & 0.921414\, -0.2196650 i \\
		\noalign{\smallskip}\hline
          \rowcolor{purple!30}
 -0.200000 & 0 & 0.530141\, -0.0288796 i & 0.611868\, -0.0288766 i & 0.693577\, -0.0288746 i & 0.775274\, -0.0288732 i & 0.856961\, -0.0288721 i \\
 -0.200000 & 1 & 0.528403\, -0.0867626 i & 0.610361\, -0.0867228 i & 0.692246\, -0.0866961 i & 0.774082\, -0.0866775 i & 0.855883\, -0.0866639 i \\
 -0.200000 & 2 & 0.524966\, -0.1450150 i & 0.607371\, -0.1448470 i & 0.689601\, -0.1447340 i & 0.771712\, -0.1446550 i & 0.853735\, -0.1445980 i \\
 -0.200000 & 3 & 0.519907\, -0.2038760 i & 0.602948\, -0.2034300 i & 0.685677\, -0.2031310 i & 0.768186\, -0.2029210 i & 0.850536\, -0.2027680 i \\
		\noalign{\smallskip}\hline
          \rowcolor{purple!30}
 -0.183333 & 0 & 0.485963\, -0.0264729 i & 0.560879\, -0.0264702 i & 0.635779\, -0.0264683 i & 0.710668\, -0.0264671 i & 0.785548\, -0.0264661 i \\
 -0.183333 & 1 & 0.484370\, -0.0795324 i & 0.559497\, -0.0794959 i & 0.634559\, -0.0794715 i & 0.709575\, -0.0794543 i & 0.784559\, -0.0794419 i \\
 -0.183333 & 2 & 0.481219\, -0.1329310 i & 0.556757\, -0.1327760 i & 0.632135\, -0.1326730 i & 0.707402\, -0.1326010 i & 0.782591\, -0.1325480 i \\
 -0.183333 & 3 & 0.476582\, -0.1868870 i & 0.552702\, -0.1864780 i & 0.628537\, -0.1862040 i & 0.704170\, -0.1860110 i & 0.779658\, -0.1858700 i \\
		\noalign{\smallskip}\hline
          \rowcolor{purple!30}
 -0.166667 & 0 & 0.441784\, -0.0240663 i & 0.509890\, -0.0240638 i & 0.577981\, -0.0240621 i & 0.646062\, -0.0240610 i & 0.714134\, -0.0240601 i \\
 -0.166667 & 1 & 0.440336\, -0.0723022 i & 0.508634\, -0.0722690 i & 0.576872\, -0.0722468 i & 0.645069\, -0.0722312 i & 0.713236\, -0.0722199 i \\
 -0.166667 & 2 & 0.437472\, -0.1208460 i & 0.506142\, -0.1207060 i & 0.574668\, -0.1206120 i & 0.643093\, -0.1205460 i & 0.711446\, -0.1204980 i \\
 -0.166667 & 3 & 0.433256\, -0.1698970 i & 0.502457\, -0.1695250 i & 0.571397\, -0.1692760 i & 0.640155\, -0.1691010 i & 0.708780\, -0.1689730 i \\
		\noalign{\smallskip}\hline
	\end{tabular}
\end{table*}


\begin{table*}
    \centering
	\caption{Massless scalar quasinormal modes obtained by using the analytic expression \eqref{wanaliticscalar}.}
	\label{tab:3}       
	\begin{tabular}{c|c|c|c|c|c|c}
		\hline\noalign{\smallskip}
		$\alpha$  & $n$ & $\ell=6$ & $\ell=7$ & $\ell=8$ & $\ell=9$ & $\ell=10$  \\
		\noalign{\smallskip}\hline\noalign{\smallskip}
        \noalign{\smallskip}\hline        
-0.333333 & 0 & 0.883561-0.048204 i & 1.019775-0.048181 i & 1.155959-0.048166 i & 1.292120-0.048155 i & 1.428267-0.048147 i \\
     \noalign{\smallskip}\hline
-0.316667 & 0 & 0.839383-0.045794 i & 0.968787-0.045772 i & 1.098161-0.045758 i & 1.227514-0.045747 i & 1.356854-0.045740 i \\
     \noalign{\smallskip}\hline
-0.300000 & 0 & 0.795204-0.043383 i & 0.917798-0.043363 i & 1.040363-0.043349 i & 1.162908-0.043340 i & 1.285440-0.043333 i \\
     \noalign{\smallskip}\hline
-0.283333 & 0 & 0.751026-0.040973 i & 0.866809-0.040954 i & 0.982565-0.040941 i & 1.098302-0.040932 i & 1.214027-0.040925 i \\
     \noalign{\smallskip}\hline
-0.266667 & 0 & 0.706848-0.038563 i & 0.815820-0.038545 i & 0.924767-0.038533 i & 1.033696-0.038524 i & 1.142614-0.038518 i \\
     \noalign{\smallskip}\hline
-0.250000 & 0 & 0.662670-0.036153 i & 0.764832-0.036136 i & 0.866969-0.036124 i & 0.969090-0.036116 i & 1.071200-0.036111 i \\
     \noalign{\smallskip}\hline
-0.233333 & 0 & 0.618492-0.033743 i & 0.713843-0.033727 i & 0.809171-0.033716 i & 0.904484-0.033709 i & 0.999787-0.033703 i \\
     \noalign{\smallskip}\hline
-0.216667 & 0 & 0.574314-0.031332 i & 0.662854-0.031318 i & 0.751373-0.031308 i & 0.839878-0.031301 i & 0.928374-0.031296 i \\
     \noalign{\smallskip}\hline
-0.200000 & 0 & 0.530136-0.028922 i & 0.611865-0.028909 i & 0.693575-0.028900 i & 0.775272-0.028893 i & 0.856960-0.028888 i \\
     \noalign{\smallskip}\hline
-0.183333 & 0 & 0.485958-0.026512 i & 0.560876-0.026500 i & 0.635777-0.026491 i & 0.710666-0.026485 i & 0.785547-0.026481 i \\
     \noalign{\smallskip}\hline
-0.166667 & 0 & 0.441780-0.024102 i & 0.509888-0.024091 i & 0.577979-0.024083 i & 0.646060-0.024078 i & 0.714134-0.024074 i \\        
       	\noalign{\smallskip}\hline
       \noalign{\smallskip}\hline
	\end{tabular}
\end{table*}


\begin{table*}
    \centering
	\caption{Relative Error for massless scalar quasinormal modes obtained by using the expressions \eqref{ErrorR}-\eqref{ErrorI} and \eqref{Error}.}
	\label{tab:5}       
    \resizebox{1.9\columnwidth}{!}{
	\begin{tabular}{c|c|c|c|c|c|c}
		\hline\noalign{\smallskip}
		$\alpha$  & $n$ & $\delta_R(\omega(\ell=6))$ & $\delta_R(\omega(\ell=7))$ & $\delta_R(\omega(\ell=8))$ & $\delta_R(\omega(\ell=9))$ & $\delta_R(\omega(\ell=10))$  \\
		\noalign{\smallskip}\hline\noalign{\smallskip}
        \noalign{\smallskip}\hline
-0.333333 & 0 & (0.000849124, 0.148071) & (0.000546469, 0.110804) & (0.000277218, 0.0866424) & (0.000233319, 0.0686883) & (0.000136867, 0.0556285) \\   
      	\noalign{\smallskip}\hline    
-0.316667 & 0 & (0.000843167, 0.148507) & (0.000469052, 0.110913) & (0.000281771, 0.0872975) & (0.000233319, 0.0681422) & (0.000111072, 0.0563933) \\
      	\noalign{\smallskip}\hline
-0.300000 & 0 & (0.000962303, 0.146690) & (0.000491990, 0.111034) & (0.000286830, 0.0857205) & (0.000233319, 0.0698412) & (0.000160205, 0.0572430) \\
       	\noalign{\smallskip}\hline
-0.283333 & 0 & (0.000962303, 0.147096) & (0.000517627, 0.111170) & (0.000292484, 0.0863984) & (0.000233319, 0.0692987) & (0.000132748, 0.0557506) \\
     	\noalign{\smallskip}\hline
-0.266667 & 0 & (0.000962303, 0.147553) & (0.000546469, 0.111322) & (0.000298845, 0.0871610) & (0.000233319, 0.0686883) & (0.000101859, 0.0566664) \\
       	\noalign{\smallskip}\hline
-0.250000 & 0 & (0.000962304, 0.148071) & (0.000448407, 0.111495) & (0.000306054, 0.0852595) & (0.000233319, 0.0679966) & (0.000160205, 0.0577043) \\
       	\noalign{\smallskip}\hline
-0.233333 & 0 & (0.000962304, 0.148662) & (0.000476425, 0.111692) & (0.000314293, 0.0860497) & (0.000233319, 0.0701706) & (0.000126865, 0.0559251) \\
       	\noalign{\smallskip}\hline
-0.216667 & 0 & (0.000962303, 0.146158) & (0.000508753, 0.111920) & (0.000323800, 0.0869616) & (0.000233319, 0.0694865) & (0.000088395, 0.0570656) \\
       	\noalign{\smallskip}\hline
-0.200000 & 0 & (0.000962303, 0.146690) & (0.000546469, 0.112186) & (0.000334890, 0.0880253) & (0.000233319, 0.0686883) & (0.000160205, 0.0549366) \\
       	\noalign{\smallskip}\hline
-0.183333 & 0 & (0.000962304, 0.147317) & (0.000591042, 0.112500) & (0.000347998, 0.0855109) & (0.000233319, 0.0677451) & (0.000117772, 0.0561947) \\
       	\noalign{\smallskip}\hline
-0.166667 & 0 & (0.000962303, 0.148071) & (0.000448408, 0.112877) & (0.000363727, 0.0866424) & (0.000233319, 0.0707635) & (0.000066852, 0.0577043) \\
       	\noalign{\smallskip}\hline
       \noalign{\smallskip}\hline
	\end{tabular}
    }
\end{table*}


\bibliographystyle{unsrt}

\bibliography{biblioIL2.bib}

\end{document}